\documentclass[conference]{IEEEtran}
\usepackage{cite} 
\usepackage{amsmath,amssymb,amsfonts,bm} 
\usepackage{algorithmic} 
\usepackage[pdftex]{graphicx} 
\usepackage{textcomp} 
\usepackage{xcolor} 
\usepackage{steinmetz}
\usepackage{tabularx}
\usepackage{array}
\usepackage{adjustbox}
\usepackage[utf8]{inputenc} 
\usepackage[T1]{fontenc} 
\usepackage{graphicx}
\usepackage{float}
\usepackage{hhline}
\usepackage{changes}
\usepackage{mathpazo} 
\usepackage{graphicx}
\usepackage{lipsum}
\usepackage{multicol}
\usepackage{blindtext}

\usepackage[autostyle=true]{csquotes} 
\definecolor{green}{rgb}{0,0.67,0}

\begin{document}
\title{Versatile and Robust Transient Stability Assessment via Instance Transfer Learning}

\author{\IEEEauthorblockN{
Seyedali~Meghdadi\IEEEauthorrefmark{1}\IEEEauthorrefmark{3},
Guido~Tack\IEEEauthorrefmark{1}, Ariel~Liebman\IEEEauthorrefmark{1},
Nicolas~Langren\'e\IEEEauthorrefmark{2}},
Christoph~Bergmeir\IEEEauthorrefmark{1}
\IEEEauthorblockA{
\IEEEauthorrefmark{1}Department of Data Science and AI,
Faculty of Information Technology, Monash University\\
\IEEEauthorrefmark{2}Data61, Commonwealth Scientific and Industrial Research Organisation (CSIRO)\\
Melbourne, Australia\\
Email: \IEEEauthorrefmark{3}seyedali.meghdadi@monash.edu}}
\maketitle
\maketitle 
\thispagestyle{plain} 
\pagestyle{plain} 
\begin{abstract} 
To support N-1 pre-fault transient stability assessment, this paper introduces a new data collection method in a data-driven algorithm incorporating the knowledge of power system dynamics. The domain knowledge on how the disturbance effect will propagate from the fault location to the rest of the network is leveraged to recognise the dominant conditions that determine the stability of a system. Accordingly, we introduce a new concept called Fault-Affected Area, which provides crucial information regarding the unstable region of operation. This information is embedded in an augmented dataset to train an ensemble model using an instance transfer learning framework. 
The test results on the IEEE 39-bus system verify that this model can accurately predict the stability of previously unseen operational scenarios while reducing the risk of false prediction of unstable instances compared to standard approaches.
\end{abstract}
\vspace{6pt}
\section{Introduction}\label{Introduction}

\IEEEPARstart{T}{he} power system is facing a range of unprecedented challenges in a new technological and economic landscape. This includes the uptake of rooftop PV and expanded adoption of renewable resources and storage systems, which are displacing large synchronous generators. The resulting changes in grid behaviour require new grid capabilities and modelling approaches, including better assessment of the ability of the system to withstand faults. In order to perform transient stability assessment (TSA) of operational instances, more extensive time-domain simulations are needed with more diverse loading conditions. The computationally intensive nature of these has led to the introduction of new data-driven machine learning (ML) approaches \cite{yousefian2017review} to accelerate TSAs. In these, data is collected from time-domain simulations, sampled from a fraction of the space of expected system states, and used off-line to train a data model. Such models, with promising results from both deep learning (DL) \cite{9105621,liu2019new} and conventional ML models \cite{6547746, ren2018multiple}, can then be used to rapidly predict the stability of other system states. 
In addition, Transfer Learning, a collection of methods for improving the performance of data models on data with different distributions or feature spaces, has recently gained attention in power system contexts. Transfer Learning for on-line TSA is proposed in \cite{8871201}, where after training a model on one fault location, the learning is transferred to assess faults at different locations. This has the advantage of requiring fewer trained models while improving accuracy.

This paper introduces a novel framework for data-driven TSA based on the understanding of a power system's dynamics. The benefit is to produce a method with versatile, robust, and interpretable outcomes for single contingency pre-fault TSA, in particular improving reliability, i.e., performance in terms of unsafe misclassifications. We achieve this by combining two datasets in a transfer learning framework: in addition to the standard dataset that captures the behaviour of the entire system, we use an auxiliary dataset that integrates the knowledge of rotor angle stability, by focusing on a small part of the network that is known to have a crucial impact on the stability for a given contingency. Moreover, we add a key element to previous approaches, emphasising the internal trade-offs in the model to minimise the cases where the model incorrectly predicts a scenario as stable.

The proposed method makes two significant improvements. Firstly, the new data collection strategy boosts the reliability of TSA predictions while maintaining a high accuracy level, regardless of the applied data model. Secondly, to avoid the low generalisation\cite{keskar1712improving}, convergence\cite{reddi1904convergence}, and interpretability of DL methods, we exploit a transfer learning-based ensemble adaptive boost model, introduced in \cite{dai2007boosting}. This also leads to high versatility (i.e. no parameter tuning is required for different contingencies) and robustness to outliers. 

We show in an empirical evaluation that this new framework outperforms standard ML and DL approaches.

\section{Methodology }\label{METHOD}
We develop a novel method that increases the reliability of data-driven TSAs by incorporating the power systems engineer's intuition and knowledge of power systems dynamics and stability. The methodology is described below in three sections.

\subsection{Fault-Affected Area and the novel data collection method}\label{FAA}
The knowledge of power system dynamics we leverage has three core pillars. Firstly, according to the equation of motion, for a given unit commitment and network configuration, variations in load result in variations in initial generator rotor angles. Secondly, after a fault, the disturbance effect will spread from the fault location to the rest of the network according to the network impedance and generator inertia \cite{sauer2017power}. Thirdly, power systems typically consist of distinct areas that are only loosely coupled.
Hence, for a given unit commitment, network configuration, and fault location, the interaction of synchronous generators in close proximity to the disturbance dominates the transient stability status.\footnote{The beyond first-swing superposition of a slow inter-area and a local plant swing mode in larger time frames \cite{Kundur2004} is not considered.}
One can empirically confirm through simulations that for a generator far from a fault location, transient instability only occurs if there is at least one unstable generator within the area where the fault occurred. This highlights the significance of variables (features in an ML sense) in close proximity to a fault location. We refer to this concept as the Fault-Affected Area (FAA).
It determines the variables that dominate transient stability, e.g.\ the output power of local synchronous generators, voltage magnitude and angle at local buses, and power flows in local transmission lines. Therefore, for any given operational scenario, the combination of these variables follows a specific pattern based on the dynamics of the FAA.
While we could train a model using only FAA scenarios, the next step combines these with scenarios sampling features from the entire network. This improves the reliability of the TSA predictions by gaining a rich feature space from which the model can better learn the dynamic behaviour of the system.
\subsection{Instance transfer learning}\label{Transductive}

Transfer learning is a technique for improving the prediction power of a model on one dataset by transforming information from a related dataset. Following this approach, we combine operational scenarios where only loads within a Fault-Affected Area vary with operational scenarios where all the loads in the network vary. The set of scenarios based on the FAA is called the \textbf{auxiliary dataset}, and it incorporates the knowledge of power system dynamics.
The set of scenarios where all loads vary is similar to the whole-system classical machine learning approach called the \textbf{same-distribution dataset}. The combination of auxiliary and same-distribution sets is called the \textbf{augmented dataset}, which is used as the training set for our models.
The trained model will be tested on a dataset that has the same underlying distribution as the same-distribution set, hence the naming. However, the probability distribution of the auxiliary set is typically different from the one of the test set and of the same-distribution set, which would prevent the successful use of classical ML approaches, where the probability distributions of the training and test sets have to be similar \cite{weiss2016survey}. Therefore, instance transfer learning is applied, a technique that was developed for situations where the distributions of data in the training and test sets are different \cite{pan2009survey}.
We use a transfer learning ensemble adaptive boost algorithm (TrAdaBoost) introduced in \cite{dai2007boosting}, where the instances are weighted so that the most useful ones have positive impact in learning the stable region of operation while the weights of the incorrectly classified samples with different distribution are adjusted to reduce their influence. 

\subsection{Internal trade-off}\label{Transductive}
The aim of this element of our framework is to reduce unsafe misclassifications to improve the reliability or safety of our method. 
Machine learning approaches often minimise overall misclassifications, i.e., both false positives ($\mathit{FP}$) and false negatives ($\mathit{FN}$). We are particularly concerned about FP, or unsafe misclassifications, which wrongly classify a scenario as stable when in reality it is unstable, since these endanger the security of the power system. A secondary objective is to achieve low safe misclassifications ($\mathit{FN}$), where the model predicts a scenario as unstable but the ground truth label is stable, so that the model does not become unnecessarily conservative. Hence we use a cost-sensitive loss function to tune the trade-off between $\mathit{FP}$ and $\mathit{FN}$.



\section{Data preparation and training procedure}\label{solutionSteps}
Our framework follows three major steps. The first is the new data collection approach, the \textit{augmented dataset}, encompassing the \textit{auxiliary} and the \textit{same-distribution} datasets. The second is the transfer learning scheme based on the TrAdaBoost model. And finally, we optimise $\mathit{FP}$ and $\mathit{FN}$ while maintaining overall accuracy. 

\begin{figure}[t]
\centerline{\includegraphics[width=7cm, height=4.6cm]{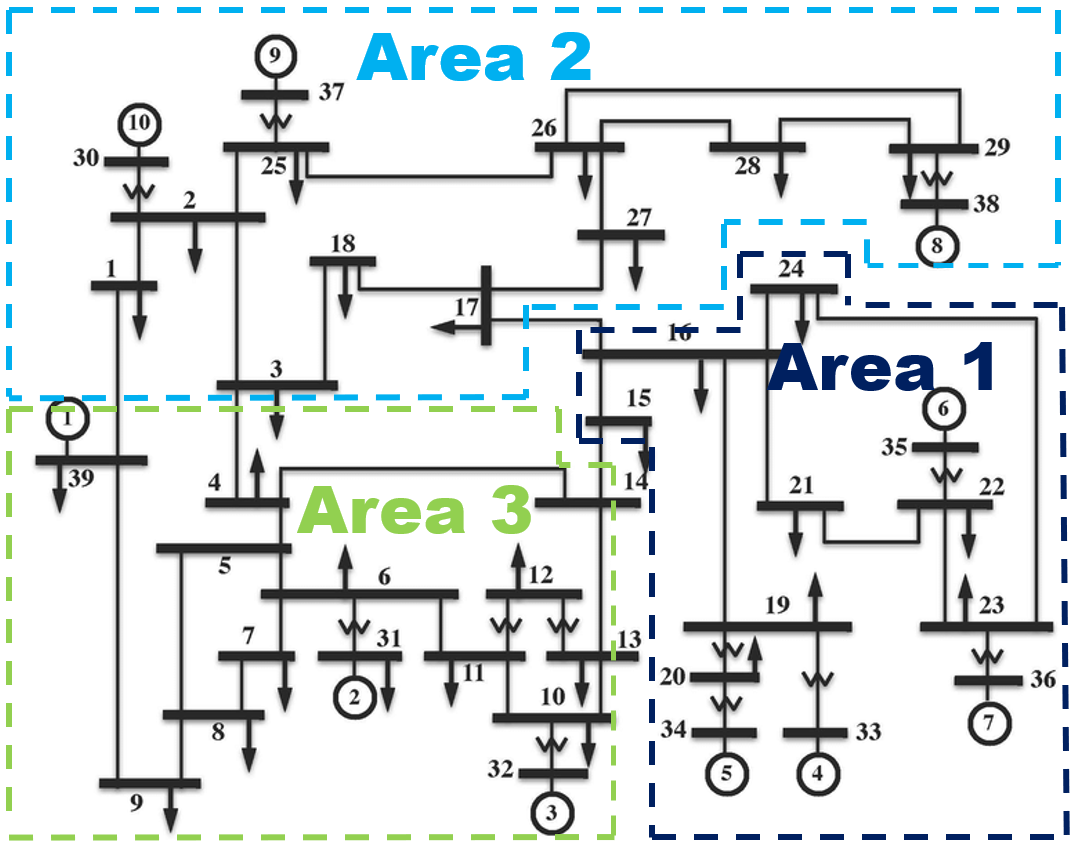}}
\caption{IEEE 39-bus system and its areas.}
\label{fig:39bus system}
\end{figure}

In order to demonstrate the performance of our new approach we use the IEEE 39-bus test system, in line with previous studies using the same experimental framework \cite{liu2019new, ren2018multiple, 6547746} to demonstrate the performance of the proposed TSA approach for a number of contingency scenarios. The network configuration along with its network areas are shown in Fig. \ref{fig:39bus system}. We approximate the impact of renewables with negative loads (i.e. large load variation ranges), while the impact of converter dynamics will be left for future work. To show that the performance of the model is independent of the fault location, we study two contingencies inside an area, line 21-22 and line 17-18, and another two on boundary lines, line 14-15 and line 3-4. Here we detail the steps to evaluate the stability for a contingency on line 21-22. 
\subsection{Used dataset}
Our aim is to map the stability of the augmented dataset, composed of the auxiliary and the same-distribution sets, to the stability of the unlabelled instances in the test set. The augmented dataset includes a \textit{same-distribution set} of 10,000 samples and an \textit{auxiliary set} of 5,000 samples, with the same proportion of classes in the two datasets.\footnote{The Sobol sampling approach is utilised throughout this study.}
To enable comparison with standard ML approaches, we also generate a set of labelled samples for training those data models. The training set includes 15,000 samples, where all loads in the system are randomly varied between 60\% and 140\% of their initial values. This large span is designed to resemble the impact of spatial and temporal uncertainties of future grids with high penetration of renewables. The trained models will be tested on a common test set of 15,000 samples with the same underlying distribution as in the training and the same-distribution sets.
 \subsubsection{Design of operational scenarios in the augmented set}\label{Distibutions}
\begin{figure}[t]
\centerline{\includegraphics[width=\columnwidth]{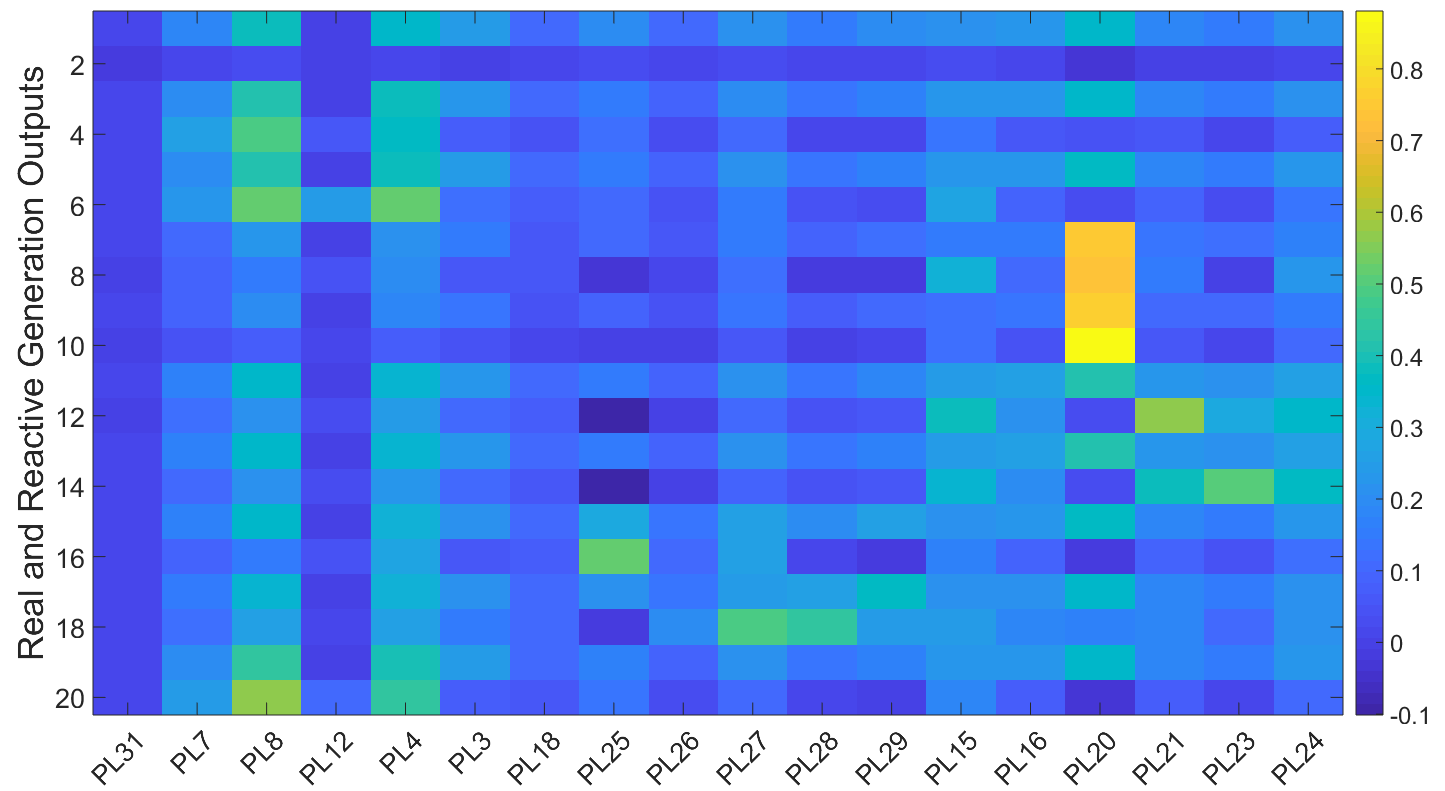}}
\caption{Sensitivity analysis: output power of generators vs. loads.}
\label{fig:sensitivity-Gen-Loads}
\end{figure}

\begin{figure}[t]
\centerline{\includegraphics[width= \columnwidth, height=5 cm]{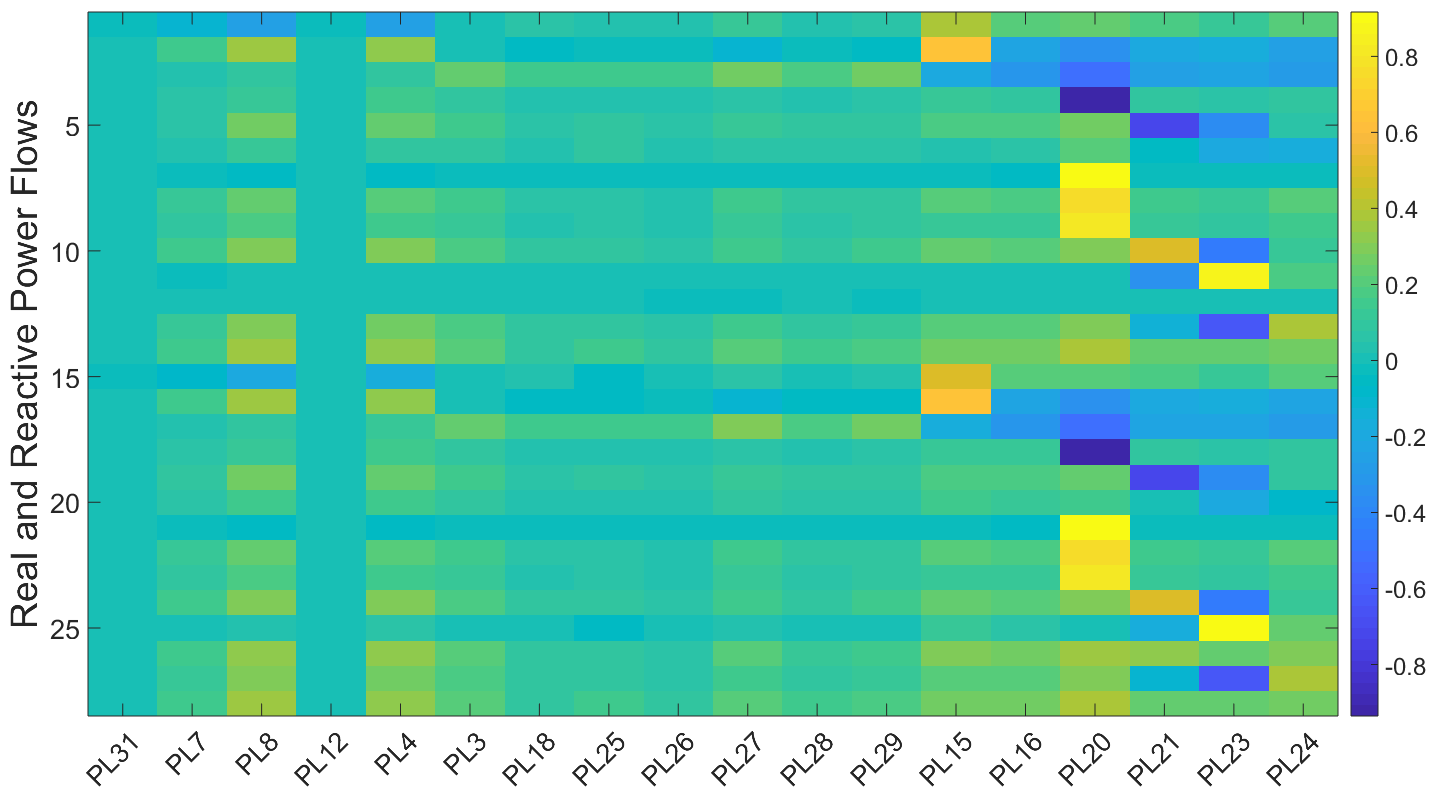}}
\caption{Sensitivity analysis: power flow of transmission lines vs. loads.}
\label{fig:sensitivity-Pflow-Loads}
\end{figure}

The same-distribution set is made by randomly varying all the loads in the system between 60\% and 140\% of initial values, which has the same underlying distribution as the test set.
On the other hand, the auxiliary set is made by randomly varying a subset of loads, as per the FAA concept, determined using two \textbf{sensitivity analyses}. The first analysis clarifies which loads in the system dominantly affect the output of local generators within the FAA.
Fig.~\ref{fig:sensitivity-Gen-Loads} depicts the correlation between load values and generator output. The highest positive correlations are shown in yellow, while the highest negative correlations are shown in dark blue. This figure reveals the great impact of the load at bus 20 and the partial importance of the loads at buses 4 and 8 in determining the output of generators. The second analysis defines which loads dominantly influence the power flow of local transmission lines.
Fig.~\ref{fig:sensitivity-Pflow-Loads} shows the correlation between load values and power flows in transmission lines, which confirms the great impact of the load at bus 20, and reveals the significance of the loads at buses 15, 16, 21, 23 and 24.
In accordance with the conclusion of the two analyses, the loads at buses 4, 8, 15, 16, 20, 21, 23 and 24 are selected to produce the load scenarios of the auxiliary set. 

\begin{figure}[b]
\centerline{\includegraphics[width= \columnwidth ]{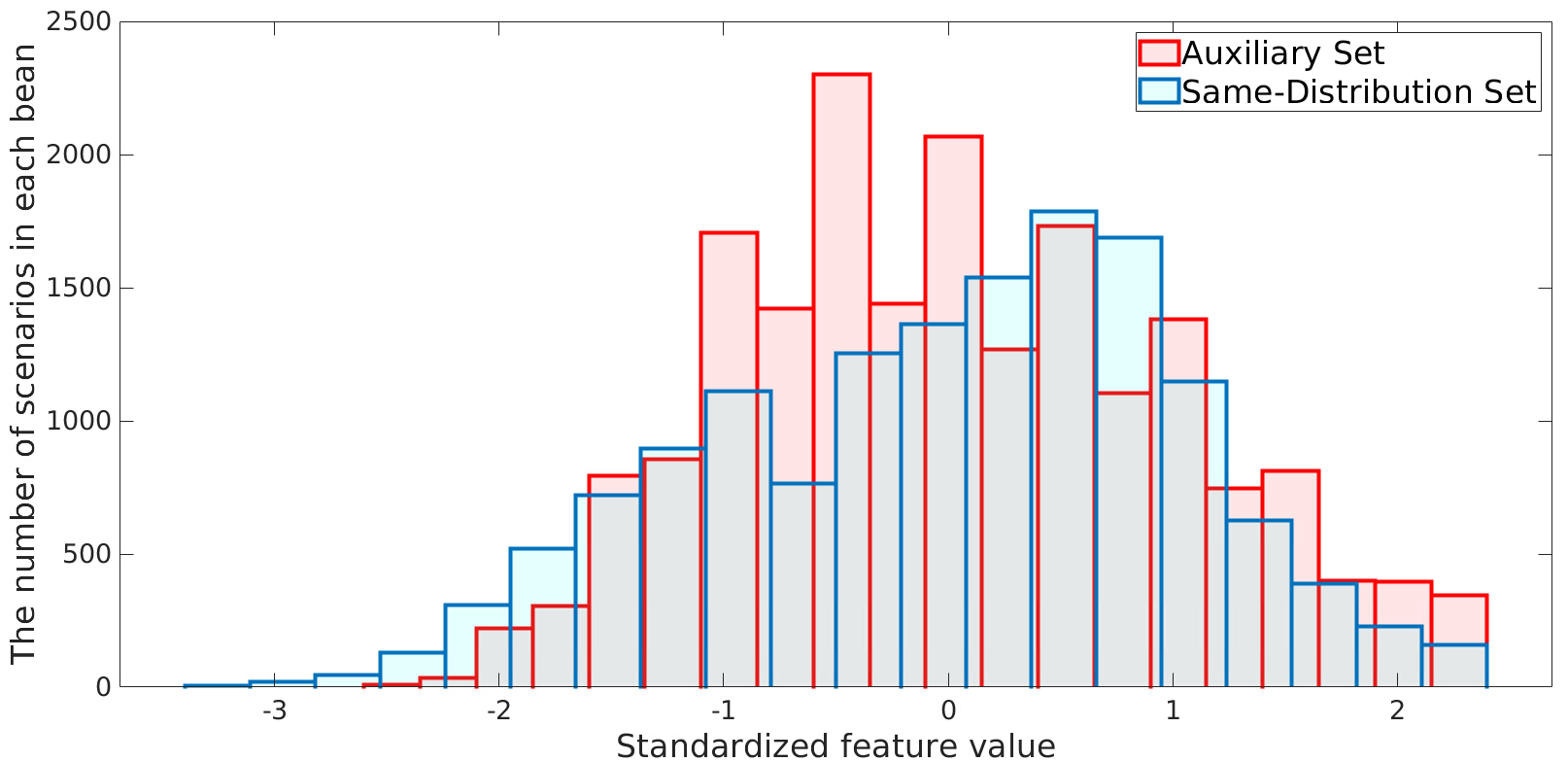}}
\caption{The range of auxiliary and same-distribution sets don't coincide}
\label{fig:FeatureUNCoverage}
\end{figure}

Moreover, since the auxiliary set and the same-distribution set include different sets of loads in their load scenarios, the distribution and range of features in the two sets are different. 
For instance, Fig.~\ref{fig:FeatureUNCoverage} shows the voltage magnitude at bus 19, where the distributions and the range of feature values in the two sets are different.
To reduce the dissimilarity in the range of feature values the selected subset of loads are varied within a 40\% to 160\% of the initial values.

\begin{figure}[t]
\includegraphics[width=\columnwidth]{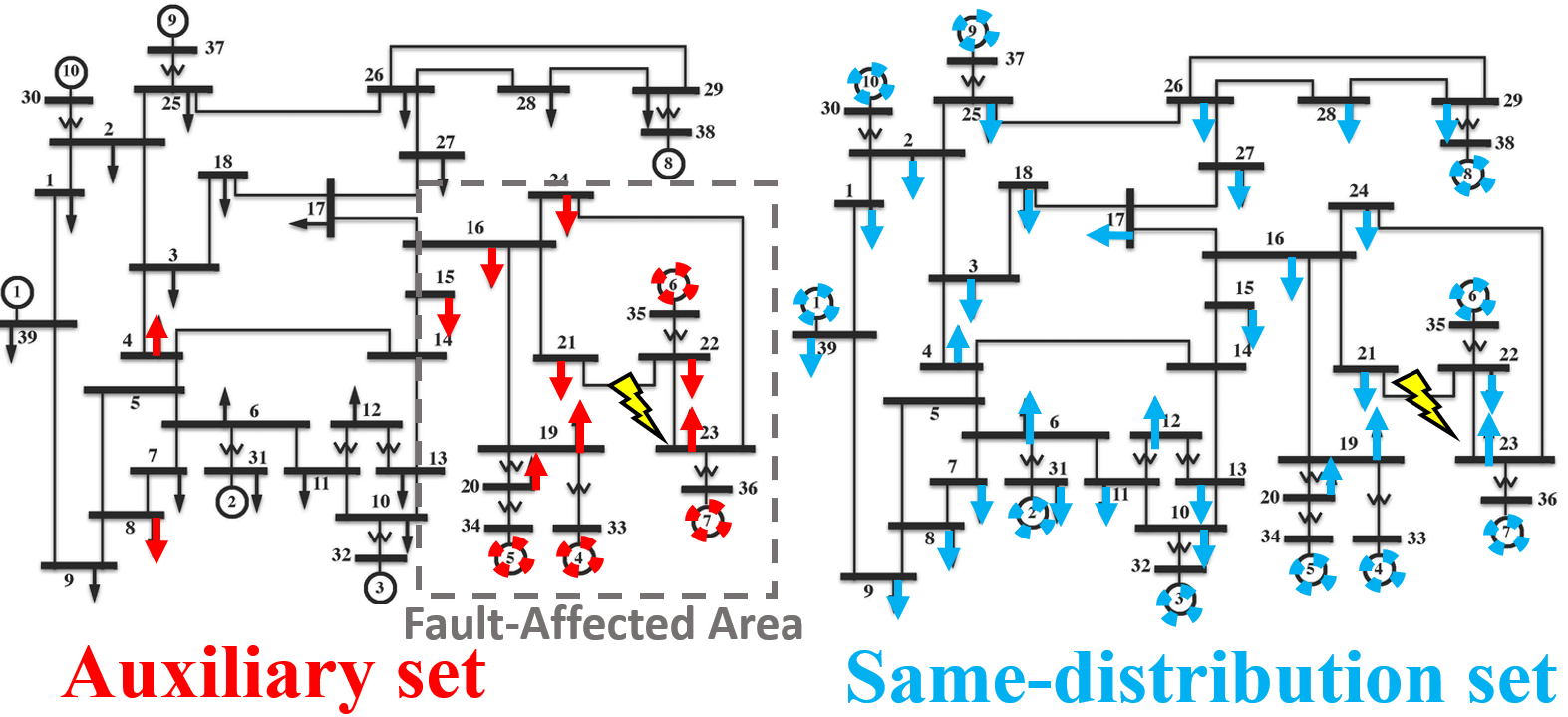} 
\caption{The schematic of the used dataset for fault on line 21-22.}
\label{fig:GraphicSourceDomain}
\end{figure}

Fig.~\ref{fig:GraphicSourceDomain} shows the overall design of operational scenarios. The auxiliary set determines the stability of generators inside the FAA (red circles) as a function of the eight varying loads (red arrows). The same-distribution set, where all the loads (blue arrows) vary, determines the stability of all the generators (blue circles).


\subsubsection{AC optimal power flow}
For every load scenario produced, the most cost-effective generation schedule is computed using an ACOPF algorithm defined by an objective function and bounded by a set of engineering laws and physical constraints \cite{zimmerman2010matpower}.

\subsubsection{Transient stability analysis} 
To determine whether each operational scenario (computed by ACOPF) remains stable after any of the contingency scenarios, a time-domain simulation is performed for 5\textit{s}, where a three-phase fault is applied at 1\textit{s} and then cleared by tripping the faulty line. The resulting answer of either \textit{stable} or \textit{unstable} is used as the classification label.
 
\subsection{Training the models}
We examine our proposed framework using the TrAdaBoost algorithm, as well as a Convolutional Neural Network (CNN) consisting of three consecutive series of convolutional and max-pooling layers followed by two fully connected layers as discussed in \cite{arteaga2019deep}. The standardised augmented set is randomly divided into training (80\%) and validation (20\%) sets while maintaining the same proportion of classes, using a stratified sampling approach, prior to training the models. 
\subsubsection{Cost-sensitive loss function}
In order to measure the performance of the models we define the cost of misclassification as the cost of classifying a point into class $X$ while its true class is $Y$. Unsafe misclassification ($\mathit{FP}$) must be minimised for the security of power system operation. Nevertheless, we are interested in keeping the safe misclassification cases ($\mathit{FN}$) as low as possible. To tune the trade-off between $\mathit{FP}$ and $\mathit{FN}$, we apply the weights $\alpha$ and $\beta$ to the misclassification costs.
To replicate the cost array of the TrAdaBoost model objective, a cost-sensitive loss function (\ref{eq:loss function}) was implemented as a customised output layer of the CNN.
\begin{equation}
\mathit{Loss} =\ -\sum_{i=1}^{N}\sum_{j=1}^{2}(t_{ij}\log y_{ij}- \alpha\times\mathit{FN} - \beta\times\mathit{FP})\
\label{eq:loss function} 
\end{equation}
Here, $N$ is the number of samples, $j$ ranges over the two classes (stable and unstable), $t_{ij}$ are the targets, $y_{ij}$ are the predicted values, and $\alpha$ and $\beta$ are the varying weights applied to the misclassification costs.
\subsubsection{Bi-objective optimisation}
Based on the parameters $\alpha$ and $\beta$, a model that results in the desirable TSA predictions can be selected. We aim to find an acceptable compromise between accuracy and unsafe misclassifications ($\mathit{FP}$), which is a bi-objective problem. We followed a \textit{grid search} approach by creating a $10\times10$ matrix of $\alpha$ and $\beta$ coefficients for both models. From the resulting $100$ options for each model, we selected the ones that minimise the $FP$ while staying within 0.2 percentage points of the maximum accuracy. We refer to this as the compromise strategy. Other trade-offs could also be achieved depending on the application, e.g. maximum accuracy or minimum $\mathit{FP}$.

\section{Results and Discussion}\label{Results}

This section presents empirical results that demonstrate the comparative advantage of using the proposed data collection approach, i.e. the \textit{augmented set}, over the existing one that uses a standard training set. We also compare the performance of TrAdaBoost with a CNN model using training and augmented sets.
To find the best combination of dataset and data model we will comprehensively analyse the results for each contingency for our proposed compromise trade-off strategy. It suggests that the best combination of model and dataset not only has a very high accuracy, but also provides the lowest unsafe misclassifications ($\mathit{FP}$), compared to the safe misclassifications ($\mathit{FN}$). Since the distribution of labels in the test set is a function of fault location, each contingency has a different label distribution. Therefore, to evaluate and compare the performance of the two data models using training and augmented datasets, we use the \textit{Recall}, \textit{Precision}, \textit{Specificity}, and \textit{Accuracy} measures reflected in Table \ref{tab:MLcomparison}.
From Table \ref{tab:MLcomparison} it firstly is evident that using the augmented set results in the lowest $\mathit{FP}$s for both models. Secondly, TrAdaBoost trained on the augmented set invariably performs best overall in terms of the compromise strategy, supported by providing the highest precision and specificity measures for all contingency scenarios. Occasionally, there is a slight trade-off between reaching the lowest $\mathit{FP}$ and the highest accuracy amongst the combinations of model and dataset. For instance, the contingency on line 21-22 shows 95.59\% accuracy for CNN trained on training set compared to 95.44\% for TrAdaBoost trained on augmented set. However, since the improvement in $FP$ (0.51\%) is much larger than the slight reduction of the accuracy (0.13\%) the trade-off is reasonably justifiable. This is also confirmed with higher precision and specificity measures.
 \begin{table}[h]
\caption{Comparative combination of models and datasets.\label{tab:MLcomparison}}
\setlength{\tabcolsep}{0.5\tabcolsep}
\begin{tabular}{||c|c|c|c|c|c|c||}
\hline
\multicolumn{7}{|c|}{Contingency on line 21-22} \\
\hline
\textbf{Model} & \textbf{DataSet} & \textbf{FP} &\textbf{Rec.} &  \textbf{Pre.}& \textbf{Spe.} & \textbf{Acc.}\\
\hline
TrAdaBoost& Augmented& 1.55\% & 92.74\%  & 96.13\% & 97.35\% & 95.44\%\\
CNN& Augmented& 1.75\% &92.48\%  & 95.63\% &  97.02\% & 95.14\%\\
TrAdaBoost& Training& 2.30\% &94.13\%  & 94.25\% & 96.16\% & 95.35\%\\
CNN& Training& 2.11\% &94.28\%  & 94.73\% & 96.47\% & 95.59\%\\
\hline
\multicolumn{7}{|c|}{Contingency on line 14-15} \\
\hline
TrAdaBoost& Augmented& 0.72\%&93.37\% & 98.33\%  & 98.68\% & 96.28\% \\
CNN& Augmented& 0.74\%&93.21\%  & 98.28\% &  98.65\%  & 96.18\%\\
TrAdaBoost& Training& 0.92\% &93.18\%  & 97.86\% & 98.32\% & 96.00\%\\
CNN& Training& 0.91\% &93.24\%  & 97.88\% & 98.34\% & 96.04\%\\
\hline
\multicolumn{7}{|c|}{Contingency on line 17-18} \\
\hline
TrAdaBoost& Augmented& 1.82\% & 89.89\% & 93.93\%  & 97.35\% &  95.01\% \\
CNN& Augmented& 1.96\% & 88.85\%  & 93.47\% &  97.14\% & 94.52\%\\
TrAdaBoost& Training& 2.24\% & 89.15\%  & 92.53\% & 96.75\% & 94.38\%\\
CNN& Training& 2.06\% & 88.53\%  & 93.13\% & 96.99\% & 94.32\%\\
\hline
\multicolumn{7}{|c|}{Contingency on line 3-4} \\
\hline
TrAdaBoost& Augmented& 1.55\%&88.12\% & 93.80\%  & 97.89\% &  95.30\% \\
CNN& Augmented& 1.65\% &87.29\%  & 93.40\% &  97.75\% & 94.95\%\\
TrAdaBoost& Training& 2.03\% &86.61\%  & 91.88\% & 97.24\% & 94.42\%\\
CNN& Training& 1.94\% &86.99\%  & 92.24\% & 97.36\% & 94.61\%\\
\hline
\end{tabular}
\end{table}

Regarding the training effort of the two models, the CNN was trained on two parallel GPUs (TESLA-V100-PCIE-16GB) while TrAdaBoost was trained on 16 parallel CPU cores (Intel-Xeon-E5-2680-v3). Although CNN was trained 20\% faster on average, it required additional engineering effort for adjusting to new datasets (i.e. different contingencies) as we faced different convergence and gradient explosion issues.
On the other hand, the TrAdaBoost took more time to train but did not require any further attention for new datasets.
This indicates the higher versatility of this model compared to CNN. 

\begin{figure}[h]
\centerline{\includegraphics[width= \columnwidth, height=5 cm]{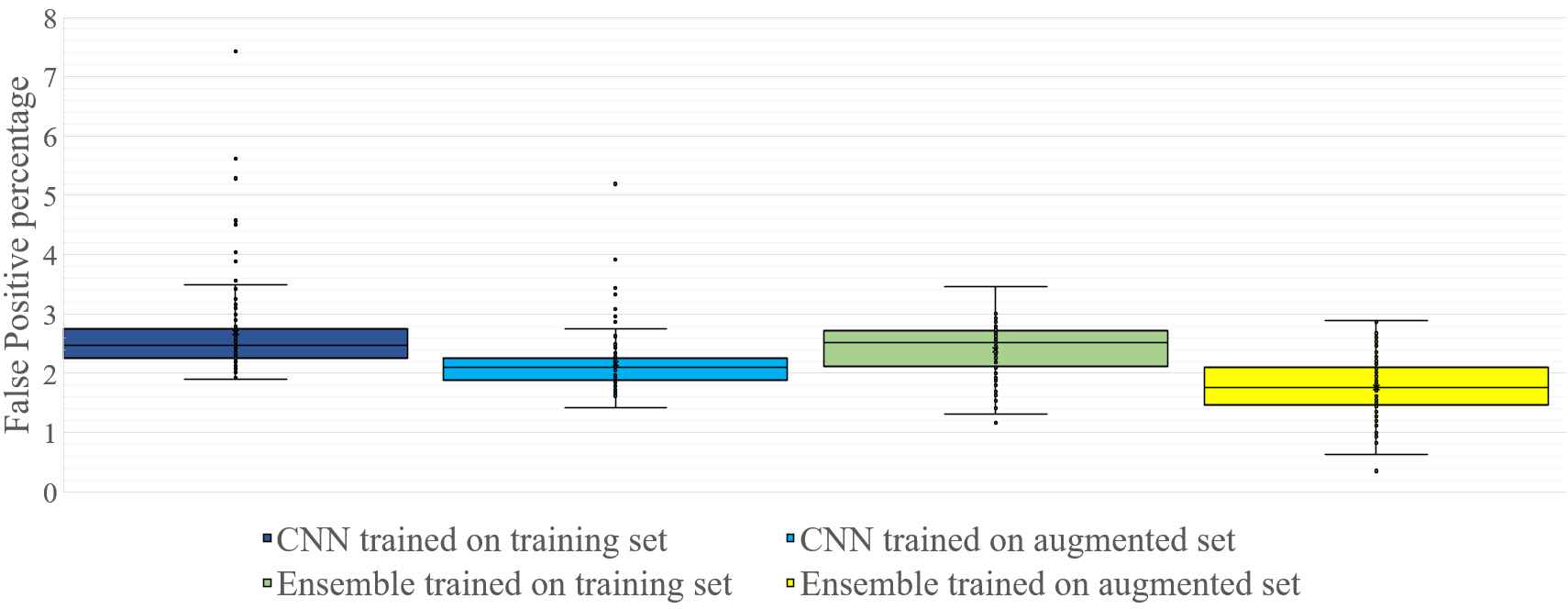}}
\caption{The False Positives of the two models and datasets.}
\label{fig:FP}
\end{figure}

Moreover, according to our experiments, CNN exhibits a larger number of outliers with high errors compared to TrAdaBoost. 
For instance, Fig. \ref{fig:FP} shows the $\mathit{FP}$ of the models and datasets for the contingency on line 21-22, proving the greater robustness of TrAdaBoost.

\section{Conclusions}\label{Conclusions}
In this paper, we explored how to improve the pre-fault prediction of transient stability of power systems using a transfer learning scheme, introducing a novel data collection approach. We incorporate an auxiliary dataset into the training process that captures crucial information about the stability of operational scenarios, focusing on the Fault Affected Area of the network.
To the best of the authors' knowledge, this is the first attempt at incorporating the knowledge of power system dynamics to develop a transfer learning mechanism for transient stability assessment. 
Using the IEEE 39-bus test case we empirically showed that the proposed data collection method achieves improved reliability compared to the standard approach, resulting in the lowest number of unsafe misclassifications on the augmented set for both models. Furthermore, the high accuracy guarantee of the TrAdaBoost model yields a high degree of confidence in the security of the analysed operational scenarios. 
Moreover, compared to deep learning models, the proposed approach provides more versatility, due to the absence of convergence and gradient explosion issues, and robustness, with fewer high-error outliers. 



\appendices
\bibliographystyle{ieeetr} 
\bibliography{AllReferences.bib}

\begin{thebibliography}{10}

\bibitem{yousefian2017review}
R.~Yousefian and S.~Kamalasadan, ``A review of neural network based machine
  learning approaches for rotor angle stability control,'' {\em
  arXiv:1701.01214}, 2017.

\bibitem{9105621}
S.~Meghdadi, G.~Tack, and A.~Liebman, ``Data-driven security assessment of the
  electric power system,'' in {\em 9th International Conference on Power and
  Energy Systems (ICPES)}, pp.~1--6, 2019.

\bibitem{liu2019new}
R.~Liu, G.~Verbi{\v{c}}, and J.~Ma, ``A new dynamic security assessment
  framework based on semi-supervised learning and data editing,'' {\em Electric
  Power Systems Research}, vol.~172, pp.~221--229, 2019.

\bibitem{6547746}
M.~{He}, J.~{Zhang}, and V.~{Vittal}, ``Robust online dynamic security
  assessment using adaptive ensemble decision-tree learning,'' {\em IEEE
  Transactions on Power Systems}, vol.~28, no.~4, pp.~4089--4098, 2013.

\bibitem{ren2018multiple}
C.~Ren, Y.~Xu, Y.~Zhang, and C.~Hu, ``A multiple randomized learning based
  ensemble model for power system dynamic security assessment,'' in {\em 2018
  IEEE Power \& Energy Society General Meeting (PESGM)}, pp.~1--5, IEEE, 2018.

\bibitem{8871201}
C.~{Ren} and Y.~{Xu}, ``Transfer learning-based power system online dynamic
  security assessment: Using one model to assess many unlearned faults,'' {\em
  IEEE Transactions on Power Systems}, vol.~35, no.~1, pp.~821--824, 2020.

\bibitem{keskar1712improving}
N.~Keskar and R.~Socher, ``Improving generalization performance by switching
  from {A}dam to {SGD},'' {\em arXiv:1712.07628}, 2017.

\bibitem{reddi1904convergence}
S.~J. Reddi, S.~Kale, and S.~Kumar, ``On the convergence of {A}dam and
  beyond,'' in {\em 6th International Conference on Learning Representations,
  {ICLR} 2018, Vancouver, BC, Canada}, 2018.

\bibitem{dai2007boosting}
W.~Dai, Q.~Yang, G.-R. Xue, and Y.~Yu, ``Boosting for transfer learning,'' in
  {\em Proceedings of the 24th international conference on Machine learning},
  pp.~193--200, 2007.

\bibitem{sauer2017power}
P.~W. Sauer, M.~A. Pai, and J.~H. Chow, {\em Power system dynamics and
  stability: with synchrophasor measurement and power system toolbox}.
\newblock John Wiley \& Sons, 2017.

\bibitem{Kundur2004}
P.~Kundur, J.~Paserba, V.~Ajjarapu, G.~Andersson, A.~Bose, C.~Canizares,
  N.~Hatziargyriou, D.~Hill, A.~Stankovic, C.~Taylor, T.~V. Cutsem, and
  V.~Vittal, ``{Definition and classification of power system stability
  IEEE/CIGRE joint task force on stability terms and definitions},'' {\em IEEE
  Transactions on Power Systems}, vol.~19, no.~3, pp.~1387--1401, 2004.

\bibitem{weiss2016survey}
K.~Weiss, T.~M. Khoshgoftaar, and D.~Wang, ``A survey of transfer learning,''
  {\em Journal of Big data}, vol.~3, no.~9, 2016.

\bibitem{pan2009survey}
S.~J. Pan and Q.~Yang, ``A survey on transfer learning,'' {\em IEEE
  Transactions on knowledge and data engineering}, vol.~22, no.~10,
  pp.~1345--1359, 2009.

\bibitem{zimmerman2010matpower}
R.~D. Zimmerman, C.~E. Murillo-S{\'a}nchez, and R.~J. Thomas, ``Matpower:
  Steady-state operations, planning, and analysis tools for power systems
  research and education,'' {\em IEEE Transactions on power systems}, vol.~26,
  no.~1, pp.~12--19, 2010.

\bibitem{arteaga2019deep}
J.-M.~H. Arteaga, F.~Hancharou, F.~Thams, and S.~Chatzivasileiadis, ``Deep
  learning for power system security assessment,'' in {\em 2019 IEEE Milan
  PowerTech}, pp.~1--6, IEEE, 2019.

\end{thebibliography}
\end{document}